\documentclass[10pt,aps,prl,showpacs,twocolumn]{revtex4-1}
\usepackage{amsmath,amssymb,graphicx}
\usepackage{mathptmx}
\usepackage{color}

\usepackage{addfont}
\addfont{OT1}{rsfs10}{\rsfs}

\usepackage{xcolor}

\usepackage{textcomp}

\usepackage{lineno}

\usepackage[scr=rsfso,cal=zapfc,frak=euler,bb=ams]{mathalfa}

\begin{document}
\title{Thermalization of orbital angular momentum beams in multimode optical fibers}

\author{E.\,V.~Podivilov$^{1,2,\dag}$}
\author{F.~Mangini$^{3,4,\dag,*}$}
\author{O.\,S.~Sidelnikov$^{1}$}
\author{M.~Ferraro$^{4}$}
\author{M.~Gervaziev$^{1}$}
\author{D.\,S.~Kharenko$^{1,2}$}
\author{M.~Zitelli$^{4}$}
\author{M.\,P.~Fedoruk$^{1}$}
\author{S.\,A.~Babin$^{1,2}$}
\author{S.~Wabnitz$^{1,4}$}

\affiliation{$^1$ Novosibirsk State University, Novosibirsk 630090, Russia}
\affiliation{$^2$ Institute of Automation and Electrometry SB RAS, 1 ac. Koptyug ave., Novosibirsk 630090, Russia}
\affiliation{$^3$ Department of Information Engineering, University of Brescia, Via Branze 38, 25123 Brescia, Italy}
\affiliation{$^4$ 
Department of Information Engineering, Electronics, and Telecommunications, Sapienza University of Rome, Via Eudossiana 18, 00184 Rome, Italy}
\affiliation{*Corresponding author: fabio.mangini@unibs.it}
\affiliation{$\dag$ These authors have contributed equally}

\date{\today}

\begin{abstract}
We present a general theory of thermalization of light in  multimode optical fibers, including optical beams with nonzero orbital angular momentum or vortex beams. A generalized Rayleigh-Jeans distribution of asymptotic mode composition is obtained, based on the conservation of the angular momentum. We confirm our predictions by numerical simulations and experiments based on holographic mode decomposition of multimode beams. This establishes new constraints for the achievement of spatial beam self-cleaning, giving previously unforeseen insights into the underlying physical mechanisms.

\end{abstract}


\maketitle

Statistical physics has been traditionally and successfully employed to describe the average properties of a large ensemble of particles, whose interactions are governed by classical mechanics. This approach lies at the basis of thermodynamics, whose laws determine the macroscopic properties of matter, that evolve in a low-dimensional or reduced phase space. Subsequently, the thermodynamic approach has been extended to describe the statistical evolution of a large number of classical electromagnetic waves, analogously to bosonic systems, such as superconductors and superfluids \cite{DYACHENKO199296,PhysRevLett.95.263901,PICOZZI20141,Sun2012,tilley2019superfluidity}. 

Peculiar is the case of multimode optical fibers (MMFs), which are an excellent study case for classical wave condensation phenomena. Indeed, Bose-Einstein condensation of the fiber modes has been demonstrated in graded-index (GRIN) MMFs \cite{PhysRevLett.125.244101}, and it can be theoretically described by a model based on a weak wave turbulence approach \cite{PRA11b}. Whereas a general model of thermalization of light in multimode systems has been recently introduced, showing that the average number of photons in each mode of the fiber obeys a Rayleigh-Jeans (RJ) distribution \cite{Wu2019}. Because of the role of high-order modes at the occurrence of thermal equilibrium in MMFs, thermalization of a multimode field is a more general situation than condensation \cite{mangini2021statistical}.

On the other hand, experimental observations have revealed that, as the input power of a laser beam coupled into a graded-index (GRIN) MMF grows above a certain threshold, the intensity speckles generated by multimode interference may spontaneously reorganize into a bell-shaped beam, which approaches the fundamental mode of the fiber \cite{KrupaPRLGPI,krupa2019multimode}. This spatial self-organization effect is known as beam self-cleaning (BSC) \cite{Krupanatphotonics}, and it has similarities with wave condensation in hydrodynamic 2D turbulent systems \cite{PhysRevLett.122.103902}. Since its first demonstration, BSC has been extensively experimentally studied
\cite{PhysRevLett.122.103902,PhysRevLett.122.123902,Deliancourt:19,
Deliancourt:19b,Leventoux:20,Fabert2020,Zitelli:21,Wu:21}, in order to fully clarify its physical mechanism. 
All of the studies reported so far in the literature agree on the fact that modal four-wave-mixing (FWM) interactions are crucial for activating BSC. 
The condensation of energy in the fundamental mode has been verified to obey the expected dependence on the initial degree of spatial correlation \cite{PhysRevLett.122.123902}, or on the internal energy of the input beam with a fixed power value \cite{PhysRevLett.125.244101}.

As a thermodynamic phenomenon, BSC can be seen as the tendency of the optical beam to experience an irreversible evolution towards a state of thermal equilibrium, which is established by conservation laws. 
Specifically, the total number of photons and the total momentum of motion must be simultaneously conserved \cite{PhysRevLett.125.244101, mangini2021statistical}.

As a matter of fact, another quantity is conserved when a beam of light propagates in waveguide systems: its total orbital angular momentum (OAM). 
First introduced by Allen et al. in 1992~\cite{allen1992orbital}, interest in OAM beams has increased tremendously, thanks to their widespread potential for applications. These range from telecommunications~\cite{wang2012terabit} 
to quantum optics~\cite{fickler2016quantum}, holography ~\cite{oraizi2020generation}
and optical tweezers~\cite{zhuang2004unraveling}.
To date, BSC has only been observed with laser beams that carry no OAM. 
In this work we extend current knowledge by describing, both theoretically and experimentally, the thermalization of OAM-carrying multimode beams in nonlinear optical fibers. We present a general theory of thermalization of light in a MMF, which takes into account the conservation of OAM. 
This permits to derive a generalized RJ distribution for the relative occupation of the fiber modes, 
which directly stems out of the conservation laws that rule FWM process in MMFs. 
Remarkably, our model shows that BSC can only be achieved by means of laser beams which do not carry any OAM. Theoretical predictions are then compared to numerical simulations, which turn out to be in excellent agreement. Finally, we carried out an experimental characterization of the thermal mode distribution at the fiber output, based on a holographic mode decomposition (MD) technique. The MD was performed by means of a phase-only spatial light modulator (SLM) \cite{Gervaziev_2020}. In our implementation, OAM is imparted to an input Gaussian laser beam by means of properly adjusting its coupling condition into the fiber \cite{mangini2020spiral}. 

In our theoretical model, we do not use the conventional Laguerre-Gauss basis for decomposing the amplitude ($A$) of a light pulse in a GRIN MMFs. Conversely, we write $A$ in terms of the real normalized \emph{radial} OAM eigenfunctions $F_{\ell,m}$, so that 
\begin{equation}\label{eigenmodes}
A(t,z, r, \phi) = \sum_{k,\ell,m} A_{\ell,m}(\omega_k,z) e^{i\omega_k t -i p_{\ell,m}(\omega_k) z + im\phi}F_{\ell,m}(r),
\end{equation}
and $2\pi \int_0^\infty rdr(F_{\ell,m}(r))^2 = 1$. 
Here, $p_{\ell,m}(\omega_k)$ and $A_{\ell,m}(\omega_k,z)$ are the propagation constant and the slowly varying amplitude of a mode with radial index $\ell$, azimuthal index $m$, and frequency $\omega_k=\omega_0+2\pi H k$, where $\omega_0$ is the carrier frequency and $H$ is the pulse repetition rate.
In the Supplementary Notes we outline the transformation from the Laguerre-Gauss basis to the OAM basis.

Although BSC is a purely spatial effect, its high threshold powers have required the use of pulsed sources for its observation. Moreover, it has been shown that BSC is accompanied by significant temporal pulse reshaping \cite{PhysRevA.97.043836}, which may be associated with a transfer of disorder between spatial and temporal degrees of freedom \cite{Laegsgaard:18}.
For this reason, in our derivation, we keep explicit the dependence of the light amplitude on frequency $\omega_k$. 

Let us normalize $|A(t,z, r, \phi)|^2$ to the beam intensity, so that for each mode $(\ell,m)$ and frequency component $\omega_k$ we may introduce the average power 
$W_{k,\ell,m}(z) = |A_{\ell,m}(\omega_k,z)|^2$, the mode energy in a pulse $E_{k,\ell,m}(z) = W_{k,\ell,m}(z)/H$, the number of photons $N_{k,\ell,m}(z) = E_{k,\ell,m}(z)/ \hbar\omega_k$, the longitudinal component of pulse momentum of motion $P_{k,\ell,m}(z) = p_{\ell,m}(\omega_k) N_{k,\ell,m}(z)$, and the OAM $M_{k,\ell,m}(z) = \hbar m N_{k,\ell,m}(z)$. 
There are four conservation laws in the FWM process. Specifically, the total energy of each pulse $E =  \sum_{k,\ell,m} E_{k,\ell,m}(z)=const $, the number of photons in each pulse 
$N =  \sum_{k,\ell,m} N_{k,\ell,m}(z)=const $, the longitudinal component of the momentum of motion of a pulse $P =  \sum_{k,\ell,m} P_{k,\ell,m}(z) =const$, and the longitudinal component of the pulse OAM $M =  \sum_{k,\ell,m} M_{k,\ell,m}(z)  =const$.
All of the conserved quantities ($E,N,P,M$) are fully defined by the injection conditions of the laser beam into the fiber. Whereas, during propagation, FWM leads to energy exchange between fiber modes, similarly to particle collisions in a gas, thus shuffling the values of $N_{\ell,m,k}(z)$. One could expect then that the photon system reaches thermodynamic equilibrium over a finite 
"time" (i.e., a finite distance z), owing to the well known Onsager's principle of detailed balance. 
When applied to mode interactions, this principle means that, at thermal equilibrium, each elementary FWM process leading to energy transfer into some modes is equally probable as its reverse process.

As it is well-known, at thermal equilibrium, which is reached after a certain propagation distance, say, $z^*$, the statistics of an ideal gas is described by the Boltzmann distribution. 
At $z>z^*$, this leads to a RJ distribution for the number of photons $N_{k,\ell,m} = \frac{T}{\hbar\omega'_{k,\ell,m} -\mu}$ occupying the mode $(k,\ell,m)$ in the coordinate system moving with the light pulse at speed $V$, and rotating with its angular velocity $\Omega$ \cite{LANDAU1980111}. Here, $T$ is the \emph{statistical} temperature of photons in a light pulse (which is analogous to the temperature of electrons in particle accelerator electron beams), $\mu$ is the chemical potential, $\omega'_{k,\ell,m}=\gamma (\omega_k-V p_{\ell,m} -\Omega m)$, and $\gamma = 1/({1-V^2/c^2})^{1/2} \simeq \sqrt{2}$ is a relativistic factor, $c$ is the speed of light in vacuum. At thermal equilibrium, the number of photons occupying mode $({\ell,m})$ with frequency $\omega_k$ in laboratory system reads as 
\begin{align}\label{Eqnave}
N_{k,\ell,m} =  \frac{T/\sqrt{2}}{\hbar\omega_{k} -\frac{\mu}{\sqrt{2}}- \hbar V p_{\ell,m} - \hbar\Omega m},
\end{align}
which is a generalized form of the RJ distribution. It is worth to noting that Eq.(\ref{Eqnave}) can be equivalently derived starting from conservation laws, without recurring to the change of coordinate system (see Supplementary Materials).


FWM scattering of waves must obey the conservation laws of $E$, $N$,$P$ and $M$, which lead to the following conditions
\begin{align}\label{Eq2}
&\omega_1+\omega_2=\omega_3+\omega_4,\\
&p_{\ell_1,m_1}(\omega_1)+p_{\ell_2,m_2}(\omega_2)\label{Eq3} = 
p_{\ell_3,m_3}(\omega_3)+p_{\ell_4,m_4}(\omega_4),\\
&m_1+m_2=m_3+m_4, \label{Eq31}
\end{align}
where $\ell_i$, $m_i$ and $\omega_i$ with $i=1,2,3,4$ characterize each of the four waves. If any of Eqs.(\ref{Eq2}), (\ref{Eq3}) or (\ref{Eq31}) are not satisfied, the scattering process is forbidden, the ergodicity hypothesis fails, and the multimode optical system never reaches its thermodynamic equilibrium.

Let us consider the FWM of narrow spectrum beams, i.e.,  
$|\omega_j -\omega_0| \ll \omega_0$. In this case, the mode propagation constants may be expanded as
$p_{\ell,m}(\omega_j) = p_{\ell,m}(\omega_0) + p_{\ell,m}'(\omega_j -\omega_0) + 0.5p_{\ell,m}'' (\omega_j -\omega_0)^2$. 
According to Eqs.(\ref{Eq2}) and (\ref{Eq3}), FWM processes within a single transverse mode have a mismatch $\delta p_{\ell,m} = p_{\ell,m}''(\omega_1 -\omega_3)(\omega_1 -\omega_4)$.
Now, the efficiency of FWM is strongly suppressed for  mismatch values larger than 
the inverse of the nonlinear length. As a result, the nonlinear spectral broadening of a light beam is restricted at long propagation distances \cite{JOSAB2007}.
When the interaction of different transverse modes is involved, the FWM mismatch may be equal to zero only occasionally, that is, for just a few quartets of waves. As a result, thermalization broadening of wave spectra fails to occur. 
Therefore, in the following we shall limit our treatment to pulses with a narrow spectrum ($\omega_k\simeq \omega$). As a matter of fact, in our experiments we use relative narrow-band picosecond pulses, whose spectral broadening is negligible over distances as long as $z^*$, which turns out to be of a few meters.



In the special case of GRIN fibers, the mode propagation constants are equidistant $p_{\ell,m} = p_{0,0} - n(2\pi/L_B)$, where  $n=2\ell+|m|$ is dubbed quantum number, whereas
$L_B$ is the self-imaging distance. 
The condition of Eqs.(\ref{Eq3}) and (\ref{Eq31}) may be met for many quartets with $n_1 +n_2 =n_3+n_4$ and $m_1 +m_2 =m_3+m_4$ simultaneously. As a result, the ergodicity hypothesis is verified, and the equilibrium  distribution (\ref{Eqnave}) is achieved after a suitable nonlinear length: it can be written as
\begin{equation}
\label{Eq5}
N_{\ell,m} = \frac{N_{0,0}}{1- (2\pi V/L_B\tilde\mu)(2\ell+|m|) + (\Omega/\tilde\mu) m},
\end{equation}
where $\tilde\mu = \mu/\hbar\sqrt{2} + V p_{0,0} - \omega$. Note that the average power of transverse mode $W_{lm}$ has the same distribution.
In the Supplementary Notes, we show that the FWM process in a multimode fiber can be described in frame of a kinetic equation approach \cite{PRA11b}, for which the distribution (\ref{Eq5}) is found to be a stationary solution. 

\begin{figure}[!h]
    \centering
\centering\includegraphics[width=8.6cm]{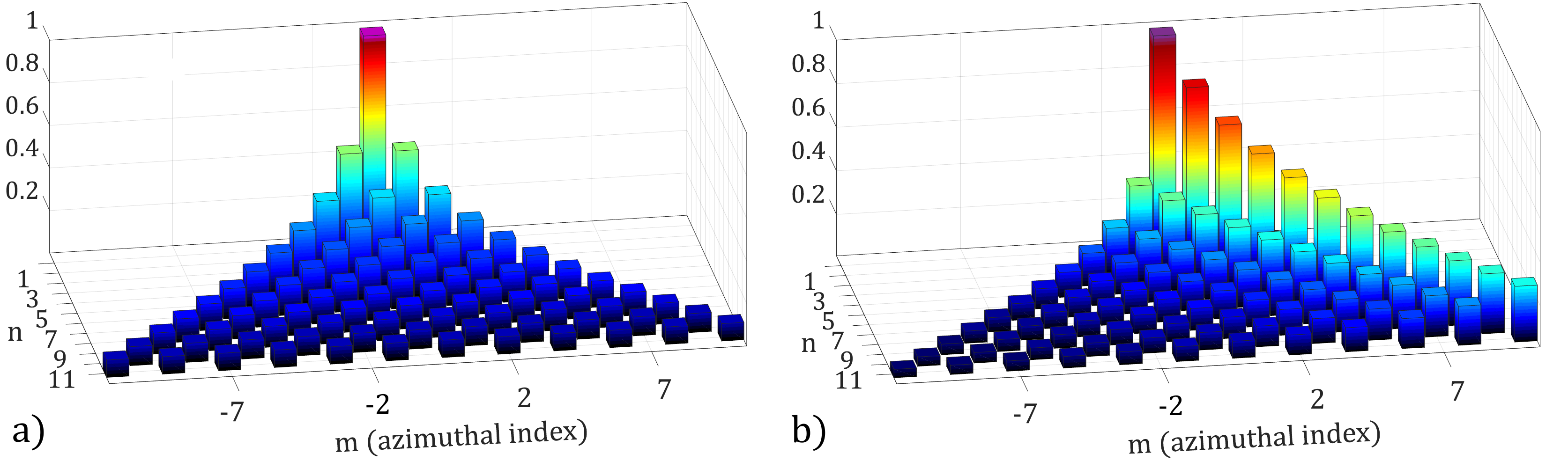}
    \caption{Equilibrium distribution for $\Omega/\tilde\mu = 0$ (a) and $\Omega/\tilde\mu = -0.75$ (b). In both plots, $N_{0,0} = 1$ and $2\pi V / L_B\tilde\mu = 1$.}
    \label{theory}
\end{figure}

Importantly, Eq.(\ref{Eq5}) shows that the equilibrium RJ distribution is asymmetric with respect to $m=0$, owing to the presence of the Lagrange's multiplier $\Omega$, which is associated to the conservation of the total OAM (see Fig.\ref{theory}b, where $\Omega > 0$). Specifically, the frequencies $(2\pi V/L_B)$ and $\Omega$ must be comparable, in order to significantly modify the symmetry of the RJ distribution around $m=0$. Whereas, if $\Omega = 0$, i.e., if the theory does not impose the conservation of the longitudinal OAM, then one recovers the conventional symmetric RJ distribution \cite{mangini2021statistical, Wu:21}, cfr. Fig.\ref{theory}a. Starting from this consideration, one can associate the presence of an OAM with the asymmetry of the mode distribution, i.e., with a non zero value of the average azimuthal index $\langle m \rangle \propto \Omega$, where
\begin{equation}
    \langle m \rangle =\sum_{\ell,m} m W_{\ell,m}.
    \label{angular-dis}
\end{equation}

The symmetry-breaking of the equilibrium distribution with respect to $m=0$ means that, whenever thermalization without condensation ($\tilde\mu \neq 0$) occurs, a  bell-shaped output beam profile cannot be obtained, unless $\langle m \rangle =\Omega = 0$, in spite of the fact that the fundamental mode is always the dominant mode in the output thermal distribution. Experimentally, this means that, in order to achieve BSC, one always need to inject a laser beam which does not carry OAM, e.g., on-axis Gaussian beams, coupled at the center of the fiber core. It is worth pointing out that BSC was earlier demonstrated by using different input laser coupling configurations. This is the case of Ref.\cite{PhysRevLett.125.244101}, where a diffuser changes the spatial distribution of the input beam, and of Ref.\cite{Fabert2020}, where fiber was tilted with respect to the laser direction. However, in both of these cases no OAM was carried by the input beam.



\begin{figure}[h!]
    \centering
\centering\includegraphics[width=8.65cm]{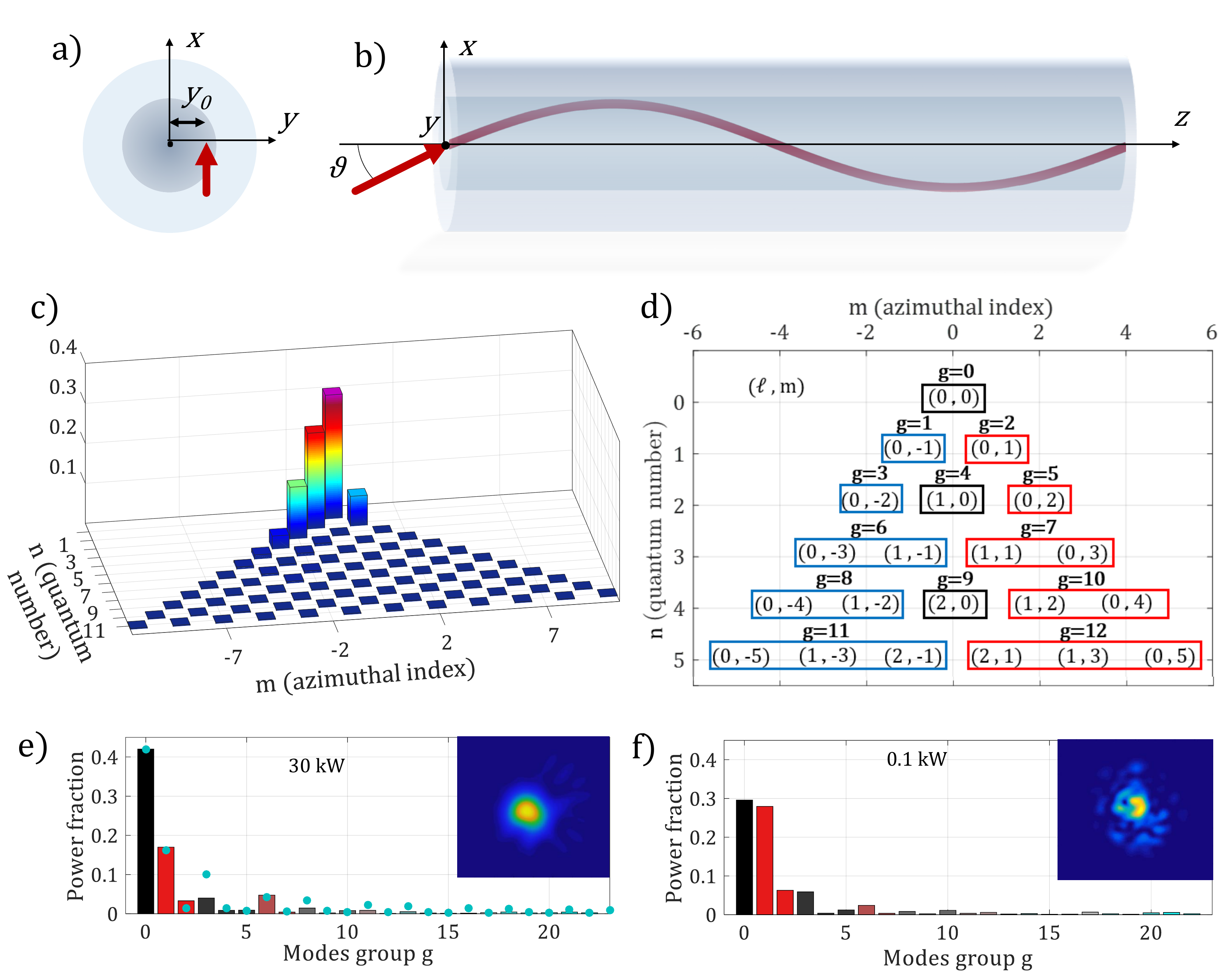}
    \caption{a,b) Sketch of front and side views of injection conditions and helical propagation carrying positive OAM. 
    c) Input mode distribution. d) Mode grouping by index $g$. e) Numerically simulated output mode distribution when $W_p = 30$ kW. Cyan dots in the 2D plot represent the values of mode power fraction, obtained by fitting experimental data with Eq.(\ref{Eq5}). 
    f) Same as e) when $W_p=0.1$ kW. Images in the inset of e) and f) represent output intensity profiles of the beams.
    }
    \label{simulations}
\end{figure}

In order to seed an OAM, we chose a peculiar input beam, i.e., a Gaussian beam which is injected with a tilt angle $\vartheta$ and a transverse offset $y_0$ with respect to the fiber axis (see Fig.\ref{simulations}a and b). Such injection condition leads to helical propagation of the laser beam inside the fiber core \cite{mangini2020spiral, mangini2022helical}: its trajectory can be visualized by the naked eye by exploiting the luminescence of fiber defects \cite{mangini2020multiphoton}. The helical trajectory carries a longitudinal OAM, which can be calculated as
\begin{equation}
    \langle m \rangle_{the} = 2\pi \frac{y_0 \sin\vartheta}{\lambda}.
    \label{angular-theo}
\end{equation}
Interestingly, the magnitude of the input OAM can be tuned by acting on the injection offset
. Specifically, the input OAM grows larger with $y_0$, and its sign can be flipped by injecting the laser at diametrically opposite points, thus reversing the helix chirality \cite{mangini2020spiral}.

The theoretical value of $\langle m \rangle$ was used to verify the validity of our numerical and experimental mode truncation. In the Supplementary Materials, we show that, as long as the offset does not exceed a few microns, there is an excellent agreement between experimental and theoretical values of $\langle m \rangle$, which are calculated numerically with Eq.(\ref{angular-dis}) or analytically via Eq.(\ref{angular-theo}).

In Fig.\ref{simulations}c) we show the mode distribution at the input of the fiber, corresponding to the injection condition $\vartheta=2^{\circ}$ and $y_0=-3$ $\mu$m. By applying Eq.(\ref{angular-dis}), this corresponds to $\langle m \rangle = 
0.75$ while $\langle n \rangle= \sum_{\ell,m}(2\ell+|m|) W_{\ell,m} = 2.23$. 

In Fig.\ref{simulations}d we display a convenient way for grouping the OAM modes, which will be used in the remainder of this work. In this way, we can emphasize the difference between groups of modes, represented by number $g$: they share the same quantum number $n$, but have different signs of $m$. Specifically, modes with odd $n$ are grouped in three blocks, one for $m<0$, one for $m>0$, and one of $m=0$. On the other hand, modes with even $n$ miss the value $m=0$, so they are grouped in two blocks only, one for $m<0$ and one for $m>0$.

In order to verify the validity of our theoretical predictions, we performed numerical simulations. Besides FWM, we also considered the effects of linear random mode coupling \cite{SIDELNIKOV2019101994}. Further details about the numerical model are reported in the Supplementary Notes. We limited our simulation to include the 78 modes with the highest values of momentum, i.e., we only considered GRIN fiber modes with $n<12$. This value represents a trade-off for maintaining the validity of the theoretical mode decomposition (the power fraction of the beam which is carried by modes with $n>11$ is negligible for our injection conditions) on the one hand, and reducing the computation time of both numerical simulations and the experimental mode reconstruction algorithm, on the other hand. 

We ran simulations for different values of the input peak power ($W_p$), in order to compare quasi-linear with highly nonlinear propagation regimes. 
In the latter ($W_p=30$ kW), the mode distribution at the output of the fiber (shown in Fig.\ref{simulations}e) turns out to be in excellent agreement with the generalized RJ distribution (\ref{Eq5}). For a clearer comparison, we report in Fig.\ref{simulations}e the experimental (histogram) and fitting (cyan dots) values of the mode power fraction. 
Whereas in the linear regime (i.e., for $W_p = 0.1 kW$), the lack of significant FWM interactions prevents mode thermalization: as a result,  a fit with Eq.(\ref{Eq5}) fails (see Fig.\ref{simulations}f). Details on simulation parameters are given in the Supplementary Notes.

\begin{figure}[ht!]
\centering\includegraphics[width=8.65cm]{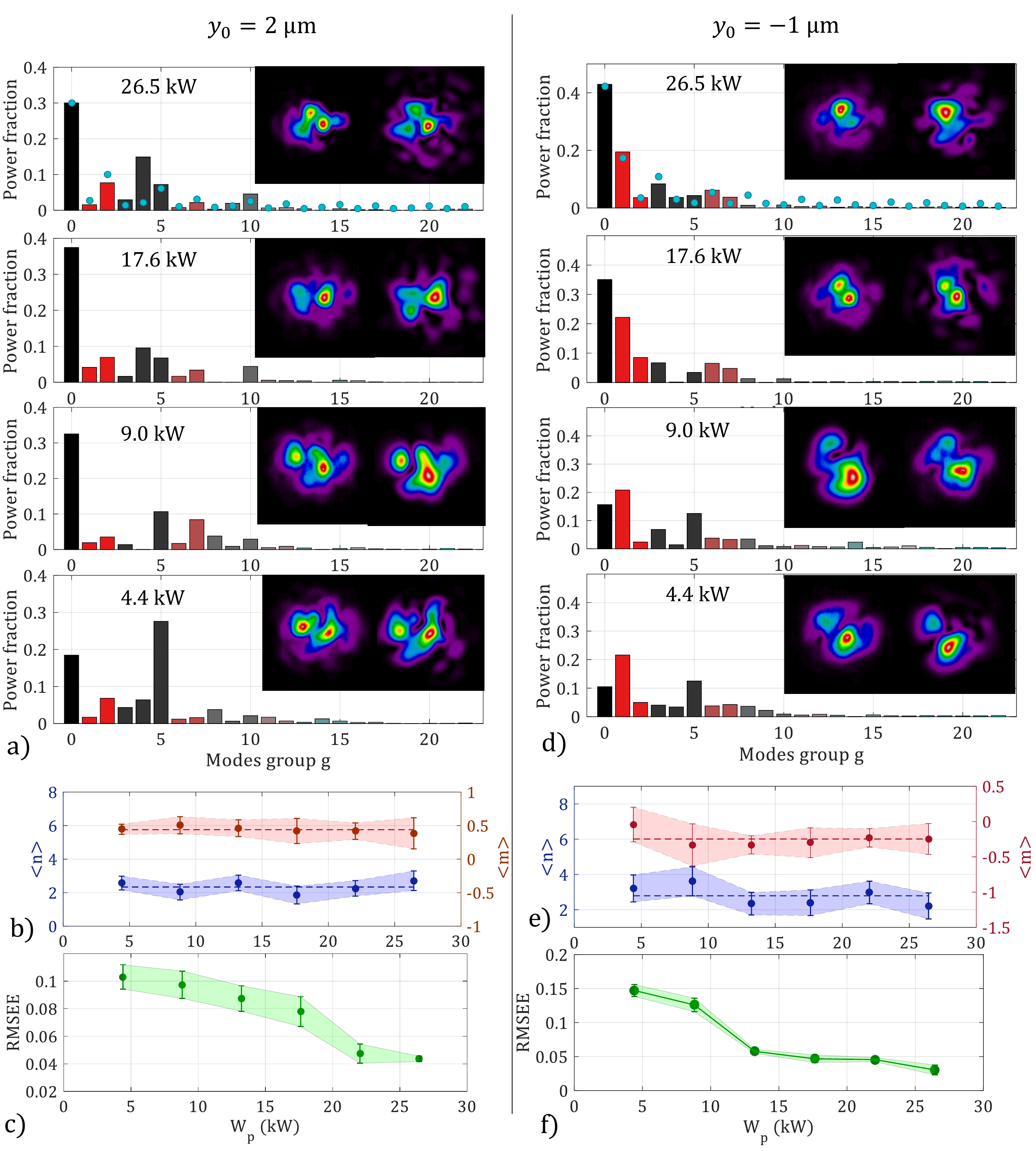}
\caption{
Experimental results. a) MD of the output beam for different values of $W_p$, when $y_0 = +2$ $\mu$m.. The inset images are the measured output beam profiles (left) and their reconstructions (right). The blue dots are extracted by fitting experimental data with Eq.(\ref{Eq5}).
b) Root-mean-square error of the experimental mode distributions with respect to the generalized RJ distribution (cyan dots in a) vs. input peak power. c) Conservation of $\langle n \rangle$ (blue) and $\langle m \rangle$ (red). The error bars are estimated by considering all of the reconstructions of the output beam near-field at each input power \cite{mangini2021statistical}. d-f) Same as a-c), when injecting the laser beam with an offset $y_0 = -1$ $\mu$m.}
\label{experiments}
\end{figure}

The validity of theory and numerics was verified by the experiments based on the MD of OAM beams at output of GRIN MMFs. We used 1 ps laser pulses at 1030 nm, and a 2 m long 50/125 GRIN MMF. A full description of the set-up is reported in the Supplementary Notes.
Here, we studied light thermalization, which is obtained by varying the input pulse peak power, with two different injection conditions. Specifically, in Fig.\ref{experiments}, we report a MD analysis of the beam output profile for input beams carrying either positive (Fig.\ref{experiments}a-c) or negative (Fig.\ref{experiments}d-f) OAM. 
This was obtained by setting $y=+2$ $\mu$m or $y=-1$ $\mu$m, respectively.
In the Supplementary Notes, we also report the limit case of $\langle m \rangle = 0$, which is achieved by injecting the laser with no offset with respect to the fiber axis.

In Fig.\ref{experiments}a and d, we report histograms of the mode power fraction of the output beams, for several values of $W_p$. As it can be seen, the mode content changes when increasing $W_p$, eventually approaching an equilibrium distribution once overcoming the critical value for thermalization. One can appreciate that the distributions at $W_p = 17.6$ kW and $W_p = 26.5$ kW are quite similar. Whereas at lower powers, significantly different output mode contents are observed.
In the inset of Fig.\ref{experiments}a and d, we compare the measured output near field intensities (images in the left column) with the MD reconstructions (images in the right column). These images are impressively similar for all input power values, thus proving the accuracy of our MD method.

The cyan dots shown in the graphs of Fig.\ref{experiments}a and d for the highest input power values provide the fitting of the experimental mode occupancy with Eq.(\ref{Eq5}). As it can be seen, a good agreement is found between the experimental mode power fractions and the prediction of the generalized RJ distribution. In Fig.\ref{experiments}b and e, we show that the root-mean-square error (RMSE) of the observed mode occupancy with respect to the equilibrium is distribution progressively reduced, when increasing $W_p$. This indicates that when enough power is provided, the FWM processes allow for reaching the ergodicity condition for the multimode system, hence its thermalization into an equilibrium distribution (\ref{Eq5}).

Finally, we proved the validity of the hypothesis behind our theoretical derivation, i.e. the conservation laws of $E$, $N$, $P$ and $M$. As a matter of fact, our MD method does not allow for estimating $N_{\ell,m}$, since only averaged quantities can be extracted from camera pictures. Nevertheless, it is well known that the energy of each pulse $E$, and accordingly the photon number $N$, are conserved in our experimental conditions, since dissipative effects, of either linear or nonlinear origin, are negligible over a few meters of fiber for picosecond pulses of few tens of kW of peak power, at wavelengths around 1 $\mu$m. 
We experimentally verified the conservation of $\langle n \rangle$  and $\langle m \rangle$, which are related to the linear and angular momentum, respectively. 
As it can be seen in Fig.\ref{experiments}c and f, when varying $W_p$, both quantities oscillate within the experimental error bars around a constant value. Specifically, we found that $\langle m \rangle \simeq 0.46$ for $y_0 = + 2$ $\mu m$ and $\langle m \rangle \simeq -0.24$ for $y_0 = - 1$ $\mu m$. This is in agreement with theoretical expectations: Eq.(\ref{angular-theo}) gives $\langle m \rangle = 0.44$ and $\langle m \rangle = -0.22$, respectively. 


In conclusion, we derived a general theoretical description of light thermalization in multimode fibers. Theoretical predictions have been confirmed by numerical and experimental studies. Remarkably, we found that the thermalization of OAM beams in GRIN MMFs may occur without the generation of bell-shaped output beams.  
Our work sheds a new light on the nonlinear dynamics and manipulation of vortex optical beams by multimode fibers.

\begin{acknowledgments}
This work was supported by the Russian Ministry of Science and Education (Grant 14.Y26.31.0017), the European Research Council (ERC) under the European Union's Horizon 2020 research and innovation programme (grant No. 740355), and Ministero dell’Istruzione, dell’Università e della Ricerca (R18SPB8227). E.P., M.G., D.K., S.B. acknowledge financial support by Russian Science Foundation (Grant 21-72-30024). 
\end{acknowledgments}

\bibliography{BIBLIO_MultimodeTemp3}

\noindent {\Huge \textbf{Supplementary \\ Notes}}




\section*{Base transformation from LG to OAM}
Since the mode decomposition code operates on a Laguerre-Gauss mode basis, while the theory is expressed in a generic OAM basis, all experimentally obtained mode decomposition data had to be transformed into an appropriate basis.
The total field in the therms of mode decomposition software is expressed as follow:
\begin{equation}\label{eq1}
U=\sum_{l,m}N_m F_{l,|m|}(A_{l,m}\cos m\phi+A_{l,-m}\sin m\phi)
\end{equation}
where $F_{l,|m|}$ the function represents Laguerre-Gaussian (LG) modes with the $(l,m)$ indexes, $A_{l,m}$ and $A_{l,-m}$ the measured values.  $N_m$ normalized coefficient that is equal to 1 for $m = 0$ and $\sqrt{2}$ for $m\neq0$ as normalization for $\sin(x)$ and $\cos(x)$ differs from $exp(ix)$. To simplify further derivations just consider $U_l$
component of the field and omit the $l$ in mode amplitudes:
\begin{equation}\label{eq2}
U_m\propto N_m(A_{m}\cos m\phi+A_{-m}\sin m\phi).
\end{equation}
Let's express the sine and cosine function by complex exponent using Euler's formula $e^{ix}=\cos x+i\sin x$ for both $m$ and $-m$ cases:
\begin{equation}\label{eq3}
\begin{cases} 
\cos m\phi=(e^{im\phi}+e^{-im\phi})/2 \\ 
\sin m\phi=(e^{im\phi}-e^{-im\phi})/2.
\end{cases}
\end{equation}
Substitute the last Eqs. (\ref{Eq3}) into (\ref{Eq2}) and let's group the coefficient before the
same exponents:
\begin{align}\label{eq4}
U_m\propto & \frac{N_m}{2}(A_{m}-iA_{-m})e^{im\phi}+\frac{N_m}{2}(A_{m}+iA_{-m})e^{-im\phi}=\nonumber\\  
= & B_m e^{im\phi}+B_{-m} e^{-im\phi}
\end{align}
where we denote $B_m$ and $B_{-m}$ as follow:
\begin{equation}\label{eq5}
\begin{cases} 
B_{l,0}=A_l \\
B_{l,m}=(A_m-iA_{-m})/\sqrt{2} \\
B_{l,-m}=(A_m+iA_{-m})/\sqrt{2}
\end{cases}
\end{equation}
In matrix form it will look like
\begin{equation}\label{eq6}
\begin{pmatrix}
B_{l,m}\\B_{l,-m}
\end{pmatrix}=\frac{1}{\sqrt{2}}
\begin{bmatrix}
1 & -i \\ 1 & i
\end{bmatrix}
\begin{pmatrix}
A_{l,m}\\A_{l,-m}.
\end{pmatrix}
\end{equation}
To calculate the angular momentum express $|B_{l,m}|^2$ through $A_{l,m}$ and $A_{l,-m}$ and take into account that $z=a+ib$ and $z-z^*=2ib$:
\begin{align}\label{eq66}
& 2|B_{l,m}|^2=|A_{l,m}|^2+|A_{l,-m}|^2+2\Im(A_{l,m}^*A_{l,-m})\\ \nonumber
& 2|B_{l,-m}|^2=|A_{l,m}|^2+|A_{l,-m}|^2-2\Im(A_{l,m}^*A_{l,-m}).
\end{align}



\section*{Derivation of the equilibrium distribution from conservation laws}

\noindent We aim at studying the thermalization of a multimode optical system taking into account the conservation laws of the number of photons $N$, the momentum $P$ and the \emph{longitudinal} orbital angular momentum (OAM), which we dub $M$. For sake of simplicity, we are considering monochromatic waves, otherwise we should have added a fourth conservation law: that of the energy of the pulse, as in the main text. 
\\
Thermalization occurs when maximizing the entropy, which reads as
\begin{equation}
    S = \ln \mathcal{W},
\end{equation}
where $\mathcal{W}$ is the number of ways in which one can distribute $N$ (indistinguishable) photons in distinct optical modes. Let us label with the index $i$ all of the $g_i$ modes having the momentum $P_i$ and OAM $M_i$. Thus the parameter $g_i$ is the mode degeneracy. Assigning to the $i$-th group of modes an occupancy $N_i$, we can write 
\begin{equation}
    \mathcal{W} = \prod_{i} \frac{(N_{i}+g_{i}-1)!}{N_{i}!(g_{i}-1)!}.
\end{equation}
Here and in the following, the index $i$ runs over all of the non-degenerate guided modes, whose (finite) number is established by the fiber cut-off. The entropy becomes:
\begin{equation}
    S = \sum_{i} (N_{i}+g_{i}-1) \ln(N_{i}+g_{i}-1) - N_{i}\ln N_{i} - \ln (g_{i}-1)!,
\end{equation}
where we used the Stirling formula $\ln x! \simeq x \ln x$ for $x \gg 1$. The conservation laws for $N$, $P$ and $M$ respectively read:
\begin{equation}
    N = \sum_i N_i,
    \label{number-i}
\end{equation}
\begin{equation}
    P = \sum_i P_i N_i,
    \label{momentum-i}
\end{equation}
and
\begin{equation}
    M = \sum_i M_i N_i.
    \label{angular-i}
\end{equation}
The conservation laws (\ref{number-i}), (\ref{momentum-i}) and (\ref{angular-i}) are the constraints for the maximization of $S$. We can introduce 3 Lagrange's multiplier $\alpha$, $\beta$ and $\gamma$ in order to find the stationary points of $S$ and we impose
\begin{equation}
    \frac{\partial}{\partial N_{i}} \left[  
    S + \alpha \bigg(N-\sum_{i} N_{i}\bigg) + \beta \bigg(P - \sum_{i} P_{i} N_{i}\bigg) + \gamma \bigg(M - \sum_{i} M_i N_{i}\bigg) \right] = 0.
\end{equation}
This gives:
\begin{equation}
    \ln\bigg(\frac{N_{i}+g_{i}-1}{N_{i}}\bigg)-\alpha-\beta P_{i} - \gamma M_i = 0
\end{equation}
\begin{equation}
    N_{i} = \frac{g_i - 1}{e^{\alpha+\beta P_i + \gamma M_i}-1},
\end{equation}
which, imposing $g_i \gg 1$, yields the Bose-Einstein distribution:
\begin{equation}
    N_{i} = \frac{g_i}{e^{\alpha+\beta P_i + \gamma M_i}-1}.
\end{equation}
Under the approximation $N_i \gg g_i$, which holds for highly multimode systems, the Bose-Eistein distribution boils down to
\begin{equation}
    N_{i} = \frac{g_i}{\alpha+\beta P_i + \gamma M_i}.
    \label{RJ-i}
\end{equation}
Note that, so far, we have extended the derivation of Ref.\cite{Wu2019} to the case of systems which conserve the longitudinal OAM. Now, we would like to apply this result to multimode fibers, where the modes are generally defined by two integer numbers ($\ell,m$). Therefore, it is convenient to get rid of the degeneracy. To do so, we need to do the following substitution:
\begin{equation}
    \sum_i g_i X_i = \sum_{\ell,m} X_{\ell,m},
    \label{i-lm}
\end{equation}
where the new sums run over all of the possible values of $\ell$ and $m$, independently of the mode degeneracy and $X$ is a generic quantity which depends on the mode indexes. Accordingly, Eq.(\ref{number-i}) becomes:
\begin{equation}
    N = \sum_{\ell,m} \frac{1}{\alpha+\beta P_{\ell,m} + \gamma M_{\ell,m}},
    \label{number-lm}
\end{equation}
and we can identify the occupancy of the mode ($\ell,m$) with
\begin{equation}
    N_{\ell,m} = \frac{1}{\alpha+\beta P_{\ell,m}+ \gamma M_{\ell,m}},
    \label{RJ}
\end{equation}
so that, by exploiting the equalities (\ref{i-lm}) and (\ref{number-lm}), the conservation laws (\ref{number-i}), (\ref{momentum-i}) and (\ref{angular-i}) respectively become
\begin{equation}
    N = \sum_{\ell,m} N_{\ell,m},
    \label{momentum}
\end{equation}
\begin{equation}
    P = \sum_{\ell,m} P_{\ell,m} N_{\ell,m},
    \label{momentumm}
\end{equation}
and
\begin{equation}
    M = \sum_{\ell,m} M_{\ell,m} N_{\ell,m}.
    \label{angular}
\end{equation}
So far, we have not chosen a basis for the mode representation. As we want to exploit the OAM of the field, it is convenient to choose a mode basis for which the \emph{longitudinal} OAM operator
\begin{equation}
    \hat{L}_z = - i \hbar \frac{\partial}{\partial\phi}
\end{equation}
is diagonalized. Such a basis is a generalization of that of Laguerre-Gauss modes, and it is dubbed OAM basis. In the latter, the radial and azimuthal variables are separated, e.g., the mode amplitude $\psi_{\ell,m}$ can be written as 
\begin{equation}
    \psi_{\ell,m}(z, r, \phi) = A_{\ell,m}(z) e^{-i P_{\ell,m} z } \cdot F_{\ell,m}(r) \cdot e^{im\phi},
\end{equation}
where $2\pi \int dr \cdot r |F(r)|^2 = 1$. Therefore, each mode $(\ell,m)$ carries a longitudinal OAM $M_{\ell,m} = m\hbar$, as 
\begin{equation}
    \hat{L}_z \cdot\psi_{\ell,m} = m\hbar\cdot \psi_{\ell,m}
\end{equation}
and the total longitudinal OAM reads as:
\begin{equation}
    M = \hbar \sum_{\ell,m} m \cdot N_{\ell,m}.
    \label{angular-m}
\end{equation}
Finally, with this basis we recover the generalized Rayleigh-Jeans distribution of our manuscript, after defining the following parameters:
\begin{equation}
    \alpha = \frac{\hbar\omega\sqrt{2}-\mu}{T},
\end{equation}
\begin{equation}
    \beta = - \frac{\hbar V\sqrt{2}}{T},
\end{equation}
\begin{equation}
    \gamma = -\frac{\hbar\Omega\sqrt{2}}{T}.
\end{equation}

\section*{Kinetic equation}

The FWM process in a multimode fiber 
can be described in frame of a kinetic equation approach \cite{PRA11b}. For simplicity of notation, let us introduce a multi-index $\mathbf{q} =(\ell,m)$.
The slowly variable mode amplitudes $A_{\mathbf{q}}(z) = A_{\ell,m}(z)$ 
obey the nonlinear evolution equation  

\begin{eqnarray}\label{E1}
&&\sum\limits_{\mathbf{q1}}F_{\mathbf{q1}}(r) e^{im_1\phi-i p_{\mathbf{q1}} z}\frac{d A_{\mathbf{q1}}}{d z}  =\nonumber \\ 
&&=-i (\omega n_2/c)
\sum\limits_{\mathbf{q2},\mathbf{q3},\mathbf{q4}} e^{-i(m_2-m_3-m_4)\phi} e^{i (p_{\mathbf{q2}} - p_{\mathbf{q3}} 
- p_{\mathbf{q4}}) z} 
\nonumber \\
&& \cdot
F_{\mathbf{q2}}(r)F_{\mathbf{q3}}(r)F_{\mathbf{q4}}(r)A^*_{\mathbf{q2}}(z)A_{\mathbf{q3}}(z) A_{\mathbf{q4}}(z)
\end{eqnarray}

\noindent where $n_2\simeq 2\cdot 10^{-16}~sm^2/W$ is the coefficient of the nonlinear 
correction to the refractive index of silica glass $\delta n_{NL}=n_2|A_{tot}|^2$,  
and $c$ vacuum speed of light. By using the orthogonality of the radial eigenfunctions, we may multiply 
Eq.(\ref{E1}) by $F_{\mathbf{q1}}(r) \exp(-i(m_1\phi - p_{\mathbf{q1}} z))$, and then integrate over the fiber cross-section. 
As a result, we obtain the propagation equation for the light amplitude in the modal representation
\begin{eqnarray}
&& \frac{d A_{\bold q1}}{d z}  = \nonumber \\
&& -i\sum\limits_{\bold q2,\bold q3,\bold q4} 
U_{\bold q1,\bold q2,\bold q3,\bold q4}
A^*_{\bold q2}(z)A_{\bold q3}(z) A_{\bold q4} (z) 
e^{i ( p_{\bold q1}+p_{\bold q2} - p_{\bold q3} - p_{\bold q4}) z } \nonumber \\
&&\cdot\Delta(m_1+m_2-m_3-m_4), \label{E2}
\end{eqnarray}
\noindent  where $\Delta(j)$ is the Kronecker's symbol, and 
$$
U_{\bold q1,\bold q2,\bold q3,\bold q4}  =
\frac{\omega n_2}{c}\int\limits_0^\infty r dr 
F_{\bold q1}(r)F_{\bold q2}(r)F_{\bold q3}(r)F_{\bold q4}(r)   
$$
\noindent is an overlap integral. 
Let us assume that, at some point $z^*$ along the fiber, the optical system reaches its steady state. Namely, all light modes have a Gaussian and delta-correlated statistics 
$\langle A_{\bold q}(z^*)A_{\bold q'}(z^*)\rangle =\Delta(l-l')\Delta(m-m') I_{\bold q}(z^*)$, and the mode power $I_{\bold q}(z)$ remains a constant for $z > z^*$. Then, exactly at $z=z^*$, one obtains 

\begin{eqnarray}\label{E3}
&& \frac{d I_{\bold q1}}{d z}|_{z=z^*}  = -2 \mbox{Re}\{\nonumber  \\
&& i\sum\limits_{\bold q2,\bold q3,\bold q4} 
U_{\bold q1,\bold q2,\bold q3,\bold q4}
\langle A^*_{\bold q1}(z^*)A^*_{\bold q2}(z^*)A_{\bold q3}(z^*) A_{\bold q4} (z^*) \rangle \nonumber \\
&& \cdot
e^{i ( p_{\bold q1}+p_{\bold q2} - p_{\bold q3} - p_{\bold q4}) z^* }\Delta(m_1+m_2-m_3-m_4))\}=  \nonumber \\
&&= 2 \mbox{Im}\sum\limits_{\bold q2} 
U_{\bold q1,\bold q2,\bold q1,\bold q2} 2I_{\bold q1}(z^*)I_{\bold q2}(z^*) =0.
\end{eqnarray}
\noindent From Eq.\ref{E2}, it follows that the amplitude of any mode $A_{\bold qj}(z)$ should change its value according to the equation
\begin{eqnarray}\label{E4}
&&\delta A_{\bold qj}(z) = A_{\bold qj}(z) - A_{\bold qj}(z^*) =\nonumber  \\
&& =- i \int\limits_{z^*}^z dz'
\sum\limits_{\bold qj',\bold qj'',\bold qj'''} 
U_{\bold qj,\bold qj',\bold qj'',\bold qj'''}
A^*_{\bold qj'}(z')A_{\bold qj''}(z') A_{\bold qj'''} (z') \nonumber \\
&& \cdot e^{i ( p_{\bold qj}+p_{\bold qj'} - p_{\bold qj''} - p_{\bold qj'''}) z' } \Delta(m_1+m_2-m_3-m_4).
\end{eqnarray}
Let us assuming, that the length $z-z^*$ is smaller than the interaction distance $L_{NL} \simeq \lambda/2\pi \delta n_{NL} $, where $\delta n_{NL}\simeq n_2 W_p/S_{eff}$ - nonlinear correction to refractive index, and $S_{eff}$ - effective area of light beam. Than Eq. (\ref{E4}) leads to a nonzero correction to the right-hand side (RHS) of Eq.(\ref{E3}), which is of the first order with respect to the small perturbation 
$\delta A$. One obtains

\begin{eqnarray}\label{E5}
&& \frac{d I_{\bold q1}}{d z}|_{z>z^*}  = -2 \mbox{Re}\{\nonumber \\
&&i \sum\limits_{\bold q2,\bold q3,\bold q4} 
U_{\bold q1,\bold q2,\bold q3,\bold q4}
\big(\langle\delta A^*_{\bold q1}(z)A^*_{\bold q2}(z^*)A_{\bold q3}(z^*) A_{\bold q4} (z^*) \rangle + \nonumber \\
&&\langle A^*_{\bold q1}(z^*)\delta A^*_{\bold q2}(z)A_{\bold q3}(z^*) A_{\bold q4} (z^*) \rangle+ \nonumber \\
&&\langle A^*_{\bold q1}(z^*)A^*_{\bold q2}(z^*)\delta A_{\bold q3}(z) A_{\bold q4} (z^*) \rangle+ \nonumber \\
&&\langle A^*_{\bold q1}(z^*)A^*_{\bold q2}(z^*)A_{\bold q3}(z^*) \delta A_{\bold q4} (z) \rangle\big)e^{i ( p_{\bold q1}+p_{\bold q2} - p_{\bold q3} - p_{\bold q4}) z^* } \nonumber \\
&&\cdot \Delta(m_1+m_2-m_3-m_4)\}.
\end{eqnarray}
By substituting here $\delta A$ from Eq.(\ref{E4}), and averaging over the Gaussian statistics, 
one obtains the general wave kinetic equation for monochromatic light in a multimode optical fiber:
\begin{eqnarray}
&& \frac{d I_{\bold q1}}{d z}|_{z>z^*}  =4 
\sum\limits_{\bold q2,\bold q3,\bold q4} 
|U_{\bold q1,\bold q2,\bold q3,\bold q4}|^2
I_{\bold q1}(z^*)I_{\bold q2}(z^*)I_{\bold q3}(z^*) I_{\bold q4} (z^*) \nonumber \\
&&
\cdot \Big(I_{\bold q1}^{-1}(z^*)+I_{\bold q2}^{-1}(z^*)-
I_{\bold q3}^{-1}(z^*) - I_{\bold q4}^{-1}(z^*)\Big) \nonumber \\ 
&& \cdot\delta_{z,z*}(p_{\bold q1}+p_{\bold q2} - p_{\bold q3} - p_{\bold q4})\Delta(m_1+m_2-m_3-m_4) \label{E6}
\end{eqnarray}
where
\begin{eqnarray}
\delta_{z,z*}(x) = \frac{\sin(x(z-z*))}{x(z-z*)}. \label{E7}
\end{eqnarray}
For a steady-state solution, the RHS of Eq.(\ref{E6}) must be equal to zero.

For GRIN multimode fibers, after averaging over the small self-imaging distance $L_B$, the function (\ref{E7}) reduces to
$\delta_{z,z*}(p_{\bold q1}+p_{\bold q2} - p_{\bold q3} - p_{\bold q4}) = 
|z-z^*|\Delta(k1+k2-k3-k4)$ for $z-z^* \gg L_B\simeq 5$~mm. 
One could check that the thermal equilibrium distribution 
\begin{eqnarray}\label{E8}
I^{-1}_{\ell,m}= I^{-1}_{0,0}(1+ a(2\ell+|m|) + b m)
\end{eqnarray} 
is indeed a steady-state solution of the wave kinetic equation (\ref{E6}): the expression in parenthesis equals to zero, thanks to the delta-symbols. The constants $I_{0,0}$, $a$ and $b$ are 
defined by the three conservation laws 
(power, momentum and angular momentum) from their input values. 
The number of photons $N_{\ell,m}$ passing through the end of the fiber at time $\tau$ has the same 
distribution, i.e., $N_{\ell,m} = \tau I_{\ell,m}/\hbar\omega$.

On the other hand, for SI multimode fibers the function (\ref{E7}) 
only reduces to the Dirac's delta-function in the limit of an infinite fiber length, i.e., whenever $z-z^* \to \infty$. Formally, one could expect to observe a thermal 
distribution in the form $I_{\ell,m}= I_{0,0}/(1+ a p_{\ell,m} + b m)$, but the mismatch
$\delta p = p_{\ell 1,m1}+p_{\ell 2,m2} - p_{\ell 3,m3} - p_{\ell 4,m4} \neq 0$ 
for any combination of transverse mode indices, so that $I_{\ell,m}$ does not reach the steady-state value if the constant $a \neq 0$ (i.e., the expression among brackets in (\ref{E7}) does not turn to zero). 
  
For small powers, the main process of energy exchange is random mode coupling.
Momentum and angular momentum conservation laws are broken by this process. Random mode coupling can be described as the scattering of the modal amplitude $A_{\ell 1,m1}$ on a noisy Bragg grating with a period $2\pi/(p_{\ell 1,m1}-p_{\ell 2,m2})$ into the modal amplitude $A_{\ell 2,m2}$, and back. 
In this case, the thermal equilibrium distribution is the equidistribution of mode powers, $I_{\ell,m}=I_{0,0}$: hence both constants $a$ and $b$ must be equal to zero.

\section*{Mode truncation}

\begin{figure}[ht!]
\centering\includegraphics[width=0.5 \textwidth]{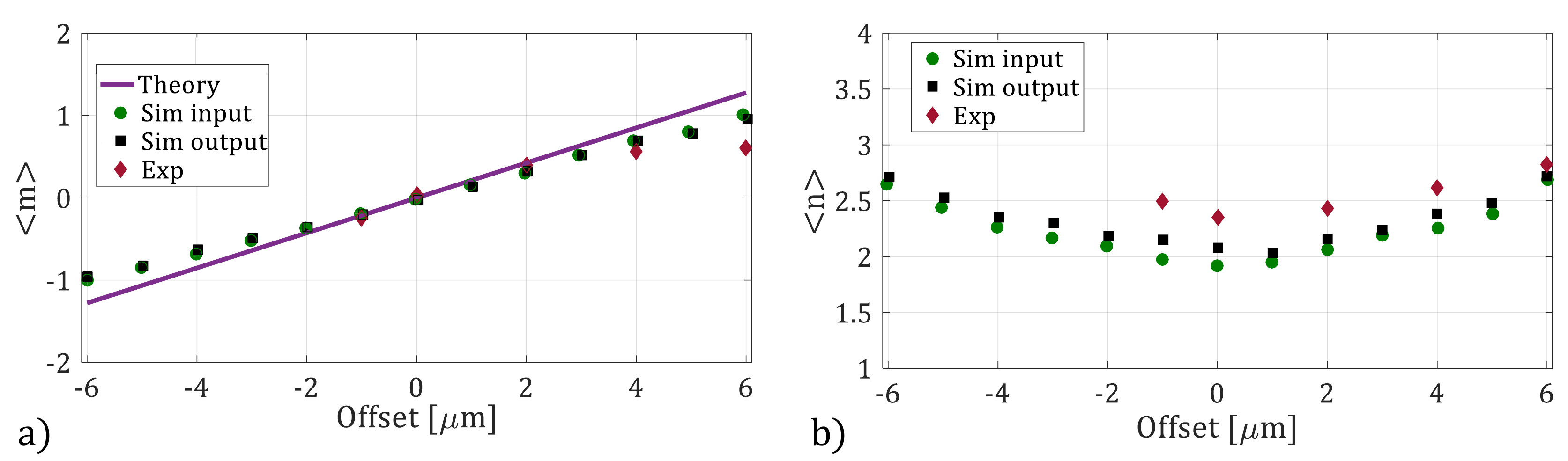}
\caption{a) Values of $\langle m \rangle$ as a function of the offset $y_0$. The solid line is calculated by Eq. (\ref{angular-theo}); Green circles, black squares and red diamonds are calculated by Eq.(\ref{angular-dis}) when considering the mode distribution at the fiber input, and at the fiber output obtained by numerical simulations and MD experiments, respectively. b) Same as a) for $\langle n \rangle$. 
}
\label{m_n}
\end{figure}

Here, we discuss the agreement between both simulation and experiments with theoretical predictions, when considering only modes with $n<12$. In Fig.\ref{m_n}, we report as green circles the values of $\langle m \rangle$ and $\langle n \rangle$ at the fiber input, when setting $\vartheta = 2^\circ$ while varying the injection offset $y_0$. The corresponding output values, calculated by numerical simulations, are shown by black squares. These are calculated by means of Eq.(\ref{angular-dis}). Similarly, the values of $\langle m \rangle$ and $\langle n \rangle$ obtained by MD experiments are shown by red diamonds. Finally, the theoretical values calculated by Eq.(\ref{angular-theo}) are shown by a solid line. As it can be seen, as long as the offset is smaller than a few microns, the MD analysis is in excellent agreement with the theory. To the contrary, when $y_0$ grows too large, a discrepancy between both numerics and experiments and the analytical prediction is found. As a consequence, in order to re-establish a coherence between those, one would need to extend the number of modes which are considered in both simulations and experiments. This would considerably increase both computation time and MD reconstruction time in the experiments. More importantly, such high-order modes may in fact not be guided in the actual GRIN fiber. Therefore, in our work we limited to offset up to 2 $\mu$m of magnitude.

\section*{Numerical methods}
To simulate the spatial evolution of the modal amplitudes of a beam in a GRIN MMF, we numerically solved the following system of coupled-mode propagation equations \cite{SIDELNIKOV2019101994}:

\begin{multline}
\frac{\partial A_{\ell,m}}{\partial z} = \sum_{\ell 1, m1} {C_{m1,m}^{\ell 1,\ell} A_{\ell 1,m1}} - \frac{\alpha}{2} A_{\ell,m} - \\ - i \frac{k n_2}{n_0}  \sum_{m1,m2,m3} \sum_{\ell 1,\ell 2,\ell 3} f^{\ell 1,\ell 2,\ell 3,\ell}_{m1,m2,m3,m} A_{\ell1,m1}^* A_{\ell 2,m2} A_{\ell 3,m3},
\label{e:cmm_stokes}
\end{multline}

where $A_{\ell,m}$ are the mode amplitudes obtained by projecting the field on the basis of Laguerre-Gaussian modes; $k = 2 \pi n_0 / \lambda$ is the wavenumber and $n_0$ is the core refractive index. The coefficients $f^{\ell 1,\ell 2,\ell 3,\ell}_{m1,m2,m3,m}$ are obtained by calculating the overlap integrals of the spatial modes distributions. The coupled-mode equations take into account the presence of the Kerr effect and linear losses ($\alpha = 2.72$ dB/km). In addition, we also included terms describing random linear coupling between all spatial modes, owing to fiber imperfections, bends and stresses. Specifically, the coefficients $C_{m,\ell}^{m1,\ell 1}$ are normally distributed random numbers with zero mean value and standard deviation $s = 2 \cdot 10^{-4}$. As done for the experimental mode decomposition, in the simulations we only consider the 78 modes whose mode number $n = 2\ell + |m| \leq 11$. Finally, in our simulations we consider a propagation distance of 10 m, whereas the spatial integration step was set to 0.01 mm.

\section*{Experimental set-up}
\begin{figure}[ht!]
\centering\includegraphics[width=8.65cm]{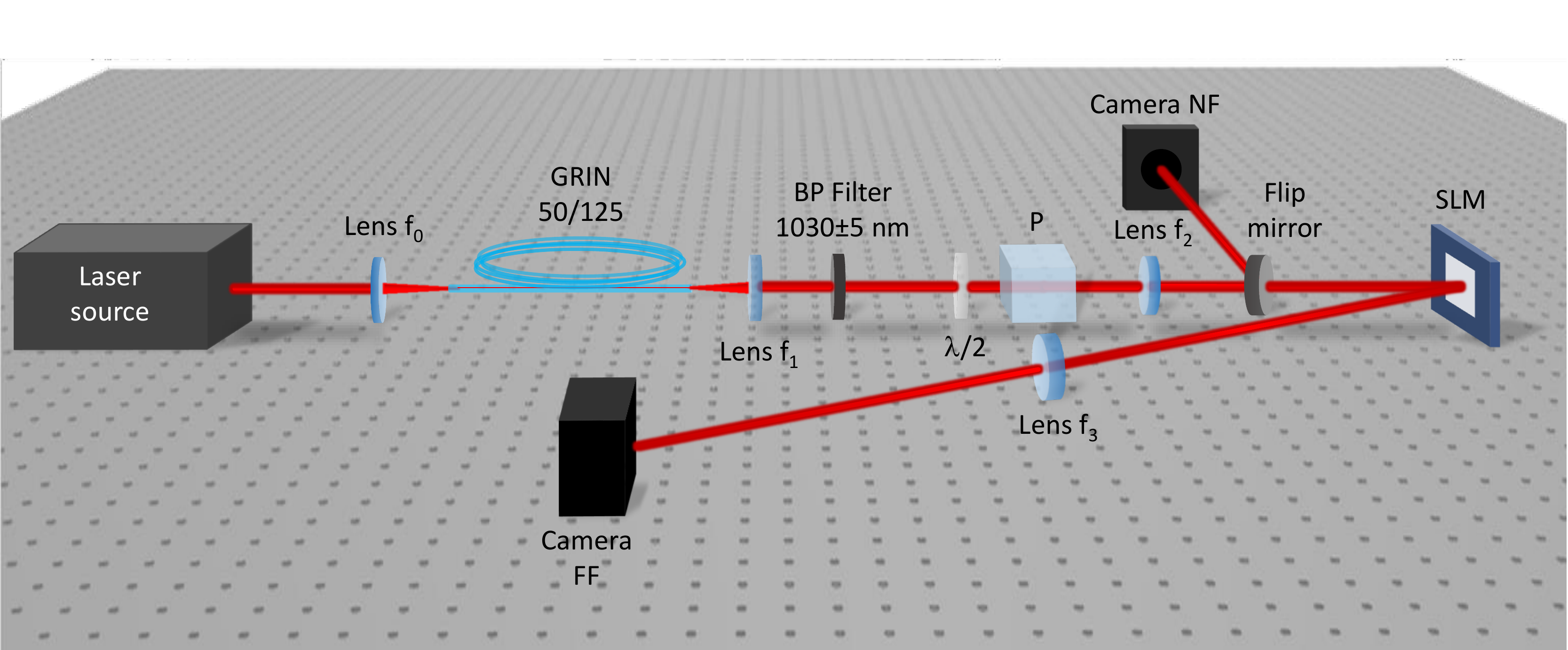}
\caption{Sketch of the experimental set-up to study the modes decomposition.}
\label{set-up}
\end{figure}

The MD experimental setup that we used to study the OAM thermalization distribution in GRIN MMFs is shown in Fig.\ref{set-up}. It consists of an ultra-short pulse laser system pumped by a femtosecond Yb-based laser (Lightconversion PHAROS-SP-HP), generating pulses with adjustable duration (by means of a dispersive pulse stretcher), at 100 kHz repetition rate and $\lambda=1030$ nm, and with Gaussian beam shape ($M^2$=1.3). The pulse shape was measured by using an autocorrelator (APE PulseCheck type 2), resulting in a sech temporal shape with pulse width equal to 1 ps. As shown in Fig.\ref{set-up}, the laser beam was injected by a positive lens ($L_0$) into the core of the GRIN fiber. The input diameter at $1/e^2$ of peak intensity was measured to be $30$ $\mu$m.
We employed 2 m long standard 50/125 GRIN fibers (GIF50E from Thorlabs), whose core radius, core refractive index along the axis, relative core-cladding index difference, numerical aperture and fundamental mode radius at $\lambda$ = 1030 nm are $r_c=25$ $\mu$m, $n_0=1.472$, $\Delta=0.0103$, $NA = 0.2$ and $r_{0,0} = 6.33$  $\mu$m, respectively. 
The fiber is positioned on a 4-Axis micro-block stages, to vary the offset and choose a proper input tilt angle $\vartheta$.
The near-field profile at the fiber output is imaged onto an SLM (Hamamatsu LCOS- X15213) by means of two confocal lenses ($L_1$, with $f_1=2.75$ mm and $L_2$, with $f_2=400$ mm). Between those, we placed a bandpass filter (BPF, $1030 \pm 5$ nm), a half-wave plate ($\lambda/2$), and a polarizer (P). Our measurement system allows us to avoid the parasitic influence of nonlinear frequency conversions, e.g., provided by Raman scattering or geometric parametric instability \cite{KrupaPRLGPI}, which is detrimental for our MD reconstruction algorithm. As a matter of fact, the phase pattern on the SLM for the profile reconstruction algorithm must be chosen at a given wavelength. We could also tune the intensity of the beam reaching the SLM by means of the $\lambda/2$.
A flip mirror (FM) is used for imaging the near field profile at the fiber output facet onto an IR camera (NF, Gentec Beamage-4M-IR). Images acquired in this way were used as a reference, in order to check the quality of the reconstruction made by the MD algorithm.
At last, a convex lens ($L_3 = 400$ mm) projects the field reflected by the SLM onto a second camera (FF camera). The lens is placed in the middle between the SLM and the camera, so that both these objects are at its focal distance. 
Finally, the beam average power both at the input and the output of the fiber was measured by a photo-diode power meter (Thorlabs).

\section*{Zero angular momentum}
\begin{figure}[!h]
\centering\includegraphics[width=0.5 \textwidth]{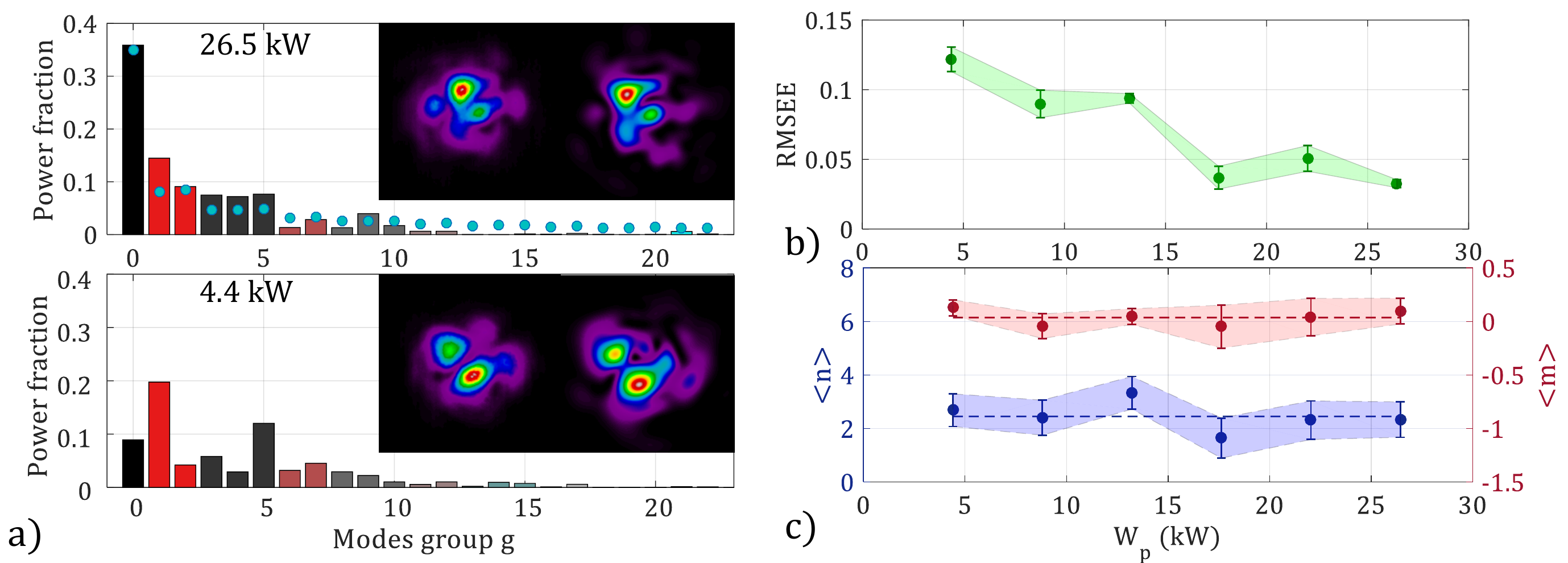}
\caption{Experimental results in the case of zero angular momentum. a) MD of the output beam when operating either in the linear (bottom panel) or in the nonlinear (top panel) regime. The inset images show the measured output beam profiles (left) and their reconstructions (right). The blue dots are extracted from the fit of the experimental data by Eq.(\ref{Eq5}).
b) Root-mean-square error of the experimental mode distributions with respect to the RJ distribution (blue dots in a) vs. input peak power. c) Conservation of $\langle n \rangle$ (blue) and $\langle m \rangle$ (red).}
\label{offset0}
\end{figure}

Here, we report the particular case of beam propagating without carrying an OAM. In this case, the distribution \ref{Eqnave} boils down to the RJ distribution. We consider the case of a meridional ray, which is achieved by injecting the laser beam without any offset with respect to the fiber axis ($y_0= 0$) while keeping a tilted geometry ($\vartheta=2^{\circ}$). In Fig.\ref{offset0}a, we report the output mode distribution when operating in the linear (bottom panel) or in the nonlinear (top panel) regime. As it can be seen, the mode content changes when varying $W_p$. Specifically, the fundamental mode becomes more populated when $W_p$ grows larger. Eventually, the output distribution reaches an equilibrium, as for the cases considered in the main text. However, in this case, when the equilibrium is reached, we may remark the symmetry of the power fraction associated with opposite signs of the azimuthal index. 
In the inset of Fig.\ref{offset0}a, we compare the measured output near field intensities (left) with the MD reconstructions (right). 
In Fig.\ref{experiments}b, we show that the root-mean-square error (RMSE) of the observed mode occupancy with respect to the RJ distribution progressively reduces, when increasing $W_p$. 
By comparing Fig.\ref{experiments}b with the experimental results in the main text, one can see that, in the abseence of OAM, thermodynamic equilibrium is reached for lower powers. 
Finally, in Fig.\ref{experiments}c we verified the conservation of $\langle m \rangle$ and $\langle n \rangle$.

\end{document}